\documentclass[12pt]{article}
\usepackage{epsf}

\def\Tr                 {\mathop{\rm Tr}}
\def\none               {\multicolumn{2}{c|}{---}}
\newcommand{\be}{\begin{equation}}
\newcommand{\ee}{\end{equation}}
\newcommand{\chinfty}{\chi^\infty}
\newcommand{\chiqu}{\chi^{\rm qu}}

\newcommand{\chiquhat}{\hat{\chi}^{\rm qu}}
\def\lsi{\raise0.3ex\hbox{$<$\kern-0.75em\raise-1.1ex\hbox{$\sim$}}}
\def\gsi{\raise0.3ex\hbox{$>$\kern-0.75em\raise-1.1ex\hbox{$\sim$}}}

\newcommand{\gsim}{\mathbin{\gsi}}
\newcommand{\MeV}{\mathop{\rm MeV}}
\newcommand{\fm}{\mathop{\rm fm}}

\begin{document}

\begin{titlepage}
\begin{flushright}
Edinburgh 2000/10 \\
OUTP-00-16P\\
hep-ph/0004180\\
\hspace{1em}\\
April 2000
\end{flushright} \begin{centering} \vfill

{\bf THE TOPOLOGICAL SUSCEPTIBILITY AND \boldmath{$f_\pi$} \\
FROM LATTICE QCD}
\vspace{0.8cm}

{\sl UKQCD Collaboration}

\vspace{0.8cm}

A. Hart$^{{\rm a}\ast}$ and 
M. Teper$^{\rm b}$

\vspace{0.3cm}
{\em $^{\rm a}$%
Department of Physics and Astronomy, University of Edinburgh,\\
Edinburgh EH9 3JZ, Scotland, UK\\}
\vspace{0.3cm}
{\em $^{\rm b}$%
Theoretical Physics, University of Oxford, 1 Keble Road, \\
Oxford OX1 3NP, England, UK\\}

\vspace{0.7cm}
{\bf Abstract.}
\end{centering}

\vspace{1ex}
\noindent
We study the topological susceptibility, $\chi$, in QCD with two quark
flavours using the lattice field configurations recently produced by
the UKQCD collaboration. We find clear evidence for the expected
suppression at small quark mass, and examine the variation of $\chi$
with this mass. The resulting estimate of the pion decay constant,
$f_{\pi} = 105 \pm 5 \ ^{+18}_{-10} \ \MeV$, is consistent with the
experimental value of $\simeq 93 \ \MeV$.  We compare $\chi$ to the
large-$N_c$ prediction and find consistency over a large range of
quark masses. The benefits of the non--perturbative action improvement
scheme and the matched lattice spacings between simulation ensembles
are discussed. We compare our results with other studies and suggest a
reason why such a quark mass dependence has not been previously seen.

\vfill

\begin{center}
$^{\ast}$ Talk presented at IoP Particle Physics 2000 conference, Edinburgh,
April 2000.
\\
\vspace{2ex}
Not submitted for publication.
\end{center}

\end{titlepage}

\section{Introduction}

The ability to access the non--perturbative sectors, and to vary
parameters fixed in Nature has made lattice Monte Carlo simulation a
valuable tool for investigating QCD and related theories, and
especially so as far as the r\^{o}le of topological excitations
is concerned.

In gluodynamics (the pure gauge or ``quenched'' theory) lattice
calculations of the continuum topological susceptibility now appear 
to be relatively free of the systematic errors arising from the 
discretisation, the finite volumes and the various measurement algorithms
employed. Attempts to measure the microscopic topological structure of
the vacuum are also well advanced (for a recent review, see
\cite{teper99}).

The inclusion of sea quarks in (``dynamical'') lattice simulations,
even at the relatively large quark masses currently employed, is
numerically extremely expensive, and can only be done for lattices
with relatively few sites (typically $16^3.32$). To avoid significant
finite volume contamination of the results, the lattice must be
relatively coarse, with a spacing $a \simeq 0.1 \ \fm$.  Compared to
quenched lattice studies at least, this is a significant fraction of
the mean instanton radius, and has so far precluded a robust, detailed
study of the local topological features of the vacuum in the presence
of sea quarks. The topological susceptibility, on the other hand, may
be calculated with some confidence and provides one of the first
opportunities to test some of the more interesting predictions for
QCD.  Indeed, it is in these measurements that we find some of the
most striking evidence for the presence of the sea quarks (or,
alternatively, for a strong quenching effect) in the lattice
simulations.

In this talk we study the topological susceptibility on 
ensembles of SU(3) lattice gauge fields that have been recently
generated by the UKQCD collaboration using a QCD lattice action
with  $N_f=2$ degenerate flavours of (fully dynamical) sea quarks.
These ensembles have been produced with two notable features.
The first is the use of an improved action, such that leading order
lattice discretisation effects are expected to depend quadratically,
rather than linearly, in the lattice spacing (just as in
gluodynamics). In addition, the action parameters have been chosen to
maintain a relatively constant lattice spacing, particularly for the
larger values of the quark mass.

We find these features allow us to see the first clear evidence for
the expected suppression of the topological susceptibility in the
chiral limit, despite our relatively large quark masses. From this
behaviour we can directly estimate the pion decay constant without 
needing to know the lattice operator renormalisation factors that
arise in more conventional calculations.

The structure of this talk is as follows. In Section~\ref{sec_theory}
we discuss the measurement of the topological susceptibility and its
expected behaviour near the chiral limit, and in the limit
of a large number of colours, $N_c$.
In Section~\ref{sec_meas} we describe the UKQCD ensembles,
and the lattice measurements of the topological susceptibility over a
range of sea quark masses. We fit these with various ans\"atze
motivated by the previous section. We compare our findings with other
recent studies in Section~\ref{sec_compare}.  Finally, we provide a
summary in Section~\ref{sec_conc}.

An early version of the results presented here can be found in
\cite{hart99},
While the results presented in this talk are still preliminary, we
do not expect any significant changes to our conclusions.
The final complete version, including estimates of the 
$\eta^{\prime}$ mass, will appear in
\cite{ukqcd_prog}.

\section{The topological susceptibility}
\label{sec_theory}

In four--dimensional Euclidean space-time, SU(3) field
configurations can be separated into classes , and moving
between different classes is not possible by a smooth
deformation of the fields.  The classes are characterised by an
integer--valued winding number.  This Pontryagin index, or topological
charge, $Q$, can be measured by integrating the local topological
charge density
\be
Q(x) = \frac{1}{2} \varepsilon_{\mu \nu \sigma \tau}
F^a_{\mu \nu}(x) F^a_{\sigma \tau}(x)
\ee
over all space time
\be
Q = \frac{1}{32 \pi^2} 
\int d^4x \cdot Q(x).
\ee
The topological susceptibility is the
expectation value of the squared charge, normalised by the volume
\be
\chi = \frac{\langle Q^2 \rangle}{V}.
\ee
Sea quarks in the vacuum induce an instanton--anti-instanton
attraction which becomes stronger as the quark masses, $m_{u}$,
$m_{d}$, \ldots, decrease towards zero (the ``chiral'' limit)
and this leads to a suppression of the topological charge and
susceptibility to leading order in the quark mass
\cite{vecchia80}
\be
\chi = \Sigma \left( \frac{1}{m_{u}} + \frac{1}{m_{d}} 
 \right)^{-1}
\ee
where 
\be
\Sigma = - \lim_{m_q \to 0} \lim_{V \to \infty} 
\langle 0 | \bar{\psi} \psi | 0 \rangle
\ee
is the chiral condensate (see 
\cite{leutwyler92} 
for a recent discussion). Here we have assumed 
$\langle 0 | \bar{\psi} \psi | 0 \rangle =
\langle 0 | \bar{u} u | 0 \rangle =
\langle 0 | \bar{d} d | 0 \rangle $
and we neglect the contribution of heavier quarks.
PCAC theory relates this to the pion decay
constant $f_\pi$ and mass $m_\pi$ as
\be
f_\pi^2 m_\pi^2 = (m_{u} + m_{d}) \Sigma + {\cal O}(m_q^2)
\ee
and we may combine these for $N_f$ degenerate light flavours 
to obtain
\be
\chi = \frac{f_\pi^2 m_\pi^2}{2 N_f} + {\cal O}(m_\pi^4)
\label{eqn_chi_pi2}
\ee
in a convention where the experimental value of $f_\pi \simeq 93 \ \MeV$.
This relation should hold in the limit $f_\pi^2 m_\pi^2 V \gg 1$. We
anticipate our results here to say that even on our most chiral
lattices the {\sc lhs} is of order 20 and so this bound is
well satisfied. Thus a calculation of $\chi$ as a function of
$m_\pi$ will allow us to obtain a value of $f_\pi$. This
method has an advantage over more conventional calculations in
that it does not require us to know difficult-to-calculate
lattice operator renormalisation constants. 

As $m_q$ and $m_\pi$ increase away from zero we expect higher order
terms to check the rate of increase of the topological susceptibility
so that, as  $m_q,m_\pi \to \infty$, it approaches the quenched value,
$\chiqu$. In fact, as we shall see below, the values of $\chi$ that
we obtain are not very much smaller than $\chiqu$. So there is the
danger of a substantial systematic error in simply applying
Eqn.~\ref{eqn_chi_pi2} at our smallest values of $m_\pi$
in order to estimate $f_\pi$. To estimate this error it
would be useful to have some understanding of how $\chi$
behaves over the whole range of $m_q$. This is the question
to which we now turn.

There are two quite different reasons why $\chi$ might not be
much smaller than $\chiqu$. The obvious first possibility
is that $m_q$ is `large'. The second possibility is more
subtle: $m_q$ may be `small' but QCD may be close to its
large-$N_c$ limit
\cite{thooft74}.
Because fermion effects are non-leading in powers of $N_c$, 
we expect $\chi \to \chiqu$ for any fixed value of $m_q$, 
however small, as the number of colours $N_c \to \infty$.
There are phenomenological reasons
\cite{thooft74,witten79}
for believing that QCD is `close' to $N_c = \infty$,
and so this is not idle speculation. Moreover in the
case of D=2+1 gauge theories it has been shown 
\cite{teper98a}
that even SU(2) is close to SU($\infty$). Cruder
calculations in four dimensions
\cite{teper98b}
indicate that the same is true there. For example
the SU(2) and SU(3) quenched susceptibilities are
very similar
\cite{teper99}.
Since our lighter quark masses straddle the strange quark mass it 
is not obvious if we should regard them as being large or small. 
We shall therefore take seriously both of the possibilities 
described above.

We start by assuming the quark mass is small but that we
are close to the large-$N_c$ limit. In this limit, the 
topological susceptibility is known 
\cite{leutwyler92} 
to vary as
\be
\chi = \frac{\chinfty m_\pi^2}
{\frac{2 N_f \chinfty}{f_\infty^2} + m_\pi^2}
\label{eqn_nlge_form}
\ee
where $\chinfty$, $f_\infty$ are the quantities at leading order in 
$N_c$. In the 
chiral limit, at fixed $N_c$, this reproduces Eqn.~\ref{eqn_chi_pi2}; 
in the large-$N_c$ limit, at fixed $m_\pi$, it plateaus to 
the quenched susceptibility, $\chinfty$, because 
$f^2_\infty \propto N_c$. The corrections to Eqn.~\ref{eqn_nlge_form}
are of higher order in $m^2_\pi$ and/or lower order in $N_c$.

We now consider the alternative possibility: that $m_q$ is not 
small, that higher order corrections to $\chi$ will be important 
for most of the values of $m_q$ at which we perform calculations,
and that we therefore need an expression for $\chi$ that
interpolates between $m_q=0$ and $m_q=\infty$. Clearly we cannot 
hope to derive such an expression, so we will simply choose one 
that we can plausibly argue is approximately correct. The
form we choose is 
\be
\chi  =  
\frac{f_{\pi}^2}{\pi N_f}  m_{\pi}^2
\arctan \left(\frac{\pi N_f}{f_{\pi}^2} \chiqu
\frac{1}{m_{\pi}^2} \right).
\label{eqn_mlge_form} 
\ee
The coefficients have been chosen so that this reproduces
Eqn.~\ref{eqn_chi_pi2} when $m_\pi \to 0$ and 
$\chi \to \chiqu + {\cal O}(1/m_\pi^4)$ when $m_\pi \to \infty$.
Thus this interpolation formula possesses the correct
limits and it approaches those limits with power-like
corrections, as we would expect.
(The precise form of the correction as
$m_\pi \to \infty$ will also depend on the physical
quantity that is used to set the overall mass scale;
we shall not attempt to determine that here.) 

We shall use the expressions in Eqns.~\ref{eqn_chi_pi2},
~\ref{eqn_nlge_form} and ~\ref{eqn_mlge_form} to
analyse the $m_q$ dependence of our calculated values 
of $\chi$ and to obtain a value of $f_\pi$ together
with an estimate of the systematic error on that
value. In addition the comparison with 
Eqn.~\ref{eqn_nlge_form} can provide us with
some evidence for whether QCD is close to its
large-$N_c$ limit or not.

\section{Lattice measurements}
\label{sec_meas}

We have calculated $\chi$ on four complete ensembles of dynamical
configurations, produced by the UKQCD collaboration
\cite{ukqcd99},
as well as on one which is still in progress. The SU(3) gauge fields
are governed by the Wilson plaquette action, with ``clover'' improved
Wilson fermions. The improvement is non--perturbative, with $c_{\rm
  sw}$ chosen to render the leading order discretisation errors
quadratic (rather than linear) in the lattice spacing, $a$.

The theory has two coupling constants. In pure gluodynamics the gauge
coupling, $\beta$, controls the lattice spacing, with larger values
reducing $a$ as we move towards the critical value at $\beta = \infty$. 
In simulations with dynamical fermions it has the same role for a fixed
fermion coupling, $\kappa$. The latter controls the quark mass, with
$\kappa \to \kappa_c$ from below corresponding to the massless
limit. In dynamical simulations, however, the fermion coupling also
affects the lattice spacing, which will become larger as $\kappa$ is
reduced (and hence $m_q$ increased) at fixed $\beta$.

The three least chiral UKQCD ensembles (by which we mean largest
$m_\pi / m_\rho$) are calculated using couplings $(\beta, \kappa) =
(5.29, 0.1340)$, $(5.26, 0.1345)$ and $(5.20, 0.1350)$ (in order of
decreasing sea quark mass). By appropriately decreasing $\beta$ as
$\kappa$ is increased, the couplings are ``matched'' to maintain a
constant lattice spacing
\cite{irving98,ukqcd99}
(which is `equivalent' to $\beta \simeq 5.9$ in gluodynamics with a
Wilson action) whilst approaching the chiral limit. Discretisation and
finite volume effects should thus be of similar sizes on these
lattices. The fourth ensemble at $(5.20, 0.1355)$ and the preliminary
results that we present for $(5.20, 0.13565)$ have lower quark masses,
but are at a reduced lattice spacing. To have maintained the matched
lattice spacing here would have required a reduction in $\beta$. As
the non--perturbative value $c_{\rm sw}(\beta)$ is not known below
$\beta=5.20$ the matching has had to be sacrificed for the continued
removal of the ${\cal O}(a)$ effects. As the lattices are all $L \gsim
1.5 \fm$, we believe that the minor reduction in our lattice volume
should not lead to significant finite volume contamination of results.

Four--dimensional lattice theories are scale free, and the
dimensionless lattice quantities must be cast in physical units
through the use of a known scale. For this work, we use the Sommer
scale
\cite{sommer94}
both to define the lattice spacing for the matching procedure, and to
set the scale. The measured value of $\hat{r}_0$ on each ensemble
corresponds to the same physical value of $r_0 = 0.49 \ \fm$.
($\hat{r}_0$ is the dimensionless lattice value of $r_0$ in lattice
units i.e. $\hat{r}_0 = r_0(a)/a$. We use the same notation for other
quantities.)  As we are in the scaling window of the theory, we can
then use the na\"{\i}ve dimensions of the various operators to relate
lattice and physical quantities, e.g.  $\hat{r}_0^4 \hat{\chi} = r_0^4
\chi + {\cal O}(a^2)$, where we have incorporated the expected
non--perturbative removal of the corrections linear in the lattice
spacing.

Further details of the parameters and scale determination are (to be)
given in
\cite{ukqcd_prog}.
Measurements were made on ensembles of 400--800 configurations of size
$L^3T = 16^3.32$, separated by ten hybrid Monte Carlo trajectories.
Correlations in the data were managed through jack--knife binning of
the data, using bin sizes large enough that neighbouring bin averages 
may be regarded as uncorrelated.

We begin, however, with a discussion of lattice operators and results
in the quenched theory.

\subsection{Lattice operators and \boldmath{$\chiqu$}}

The simplest lattice topological charge density operator is
\be
\hat{Q}(n) = \frac{1}{2} \varepsilon_{\mu \nu \sigma \tau}
\Tr U_{\mu \nu}(n) U_{\sigma \tau}(x)
\label{eqn_qlat_op} 
\ee
where $U_{\mu \nu}(n)$ denotes the product of SU(3) link variables
around a given plaquette. We use a reflection-symmetrised 
version and form
\begin {eqnarray}
\hat{Q} & = & \frac{1}{32\pi^2} \sum_n \hat{Q}(n) \\
\hat{\chi} & = & \frac{\langle \hat{Q}^2 \rangle}{L^3T}
\end{eqnarray}
with $L^3T$ the lattice volume.

In general, $\hat{Q}$ will not give an integer--valued topological
charge due to finite lattice spacing effects.
There are at least three sources of these. First is the breaking of
scale invariance by the lattice which leads  to the smallest 
instantons having a suppressed action (at least with
the Wilson action) and a topological charge less than unity
(at least with the operator in Eqn.~\ref{eqn_qlat_op}). 
We do not address this problem in this study, 
although attempts can be made to correct for it
\cite{smith98},
but simply accept this as part of the overall ${\cal O}(a^2)$ error.
In addition to this, the underlying topological 
signal on the lattice is distorted 
by the presence of large amounts of UV noise on the scale of
the lattice spacing
\cite{divecchia81},
and by a multiplicative renormalisation factor
\cite{pisa88}
that is unity in the continuum, but otherwise suppresses the observed
charge. Various solutions to these problems exist
\cite{teper99}.
In this study we opt for the `cooling' approach. Cooling explicitly
erases the ultraviolet fluctuations so that the perturbative lattice
renormalisation factors for the topological charge and susceptibility
are driven to their trivial continuum values, leaving ${\cal O}(a^2)$
corrections that may be absorbed into all the other lattice
corrections of this order. We cool by moving through the lattice in a
``staggered'' fashion, cooling each link by minimising the Wilson
gauge action applied to each of the three SU(2) subgroups in the link
element in turn. (The Wilson gauge action is the most local, and thus
the most efficient at removing short distance fluctuations whilst
preserving the long range correlations in the fields.)  Carrying out
this procedure once on every link constitutes a cooling sweep (or
``cool''). The violation of the instanton scale invariance on the
lattice, with a Wilson action, is such that an isolated instanton
cooled in this way will slowly shrink, and will eventually disappear
when its core size is of the order of a lattice spacing, leading to a
corresponding jump in the topological charge. Such events can, of
course, be detected by monitoring $\hat{Q}$ as a function of the
number of cooling sweeps, $n_c$. Instanton--anti-instanton pairs may
also annihilate, but this has no net effect on $\hat{Q}$. All this
does, however, motivate us to perform the minimum number of cools
necessary to obtain an estimate of $\hat{Q}$ that is stable with
further increasing $n_c$ (subject to the above). In our calibration
studies we found $\langle \hat{Q} \rangle(n_c)$ and $\langle \hat{Q}^2
\rangle(n_c)$ to be stable within statistical errors for $n_c \gsim
5$, and the results presented here are for $n_c = 10$.

We also remark that on a lattice one obviously loses instantons with
sizes $\rho \leq {\cal O}(a)$. Since the (pure gauge) instanton
density decreases as $\rho^6$ this would appear to induce a negligible
${\cal O}(a^7)$ error in the susceptibility. It can be numerically
substantial, however, for the coarse lattices often used in dynamical
simulations.

In general, then, we expect the topological charge and susceptibility
to be suppressed at non--zero lattice spacing. In gluodynamics
with the Wilson action this can be fitted well by just a leading
order, and negative, correction term
\cite{teper99}
\be
\hat{r}_0^4 \hat{\chi} = 0.072 - 0.208/\hat{r}_0^2
\label{eqn_chiqu_int}
\ee 
in the window of lattice spacings that is comparable to those used 
in current dynamical simulations. We shall refer to this formula 
in what follows.

\subsection{Sea quark effects in the topological susceptibility}

\begin{figure}[tb]
\begin{center}
\leavevmode
\epsfysize=510pt
\epsfbox[20 30 620 730]{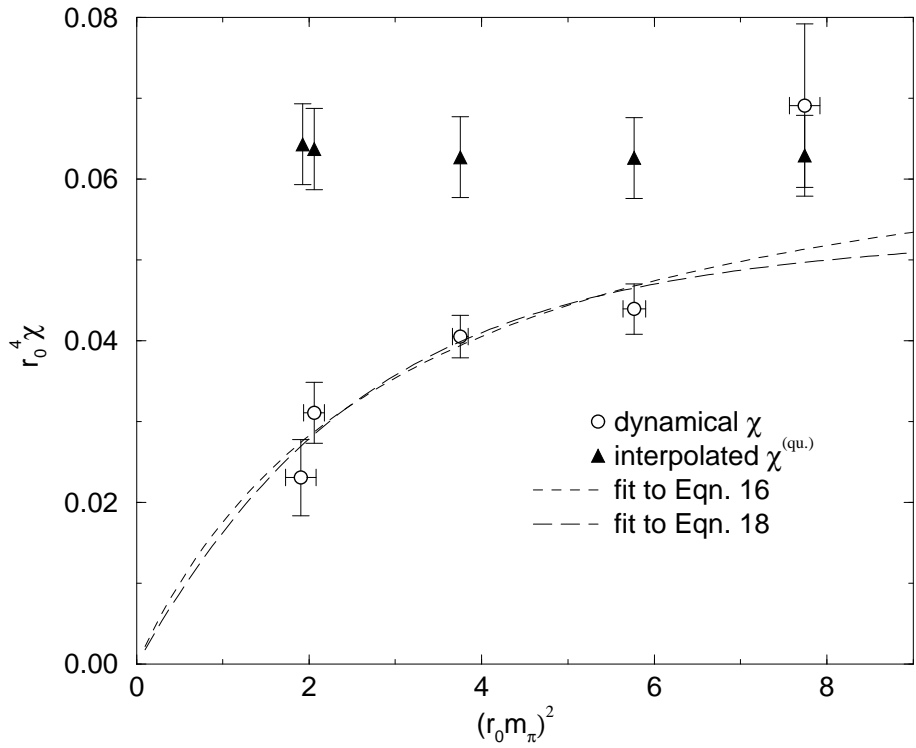}
\end{center}
\vspace{-1.0cm}
\caption[]{\label{fig_r04}
  {\it The measured topological susceptibility, with interpolated
    quenched points at the same $\hat{r}_0$ and fits independent of
    the quenched points.}}
\end{figure}

In Table~\ref{tab_chi} we give our estimates of the topological
susceptibility. We now convert these raw lattice values into
physical units using $r_0$ as the physical scale. In 
Fig.~\ref{fig_r04} we plot the resulting values of 
$\hat{r}_0^4\hat{\chi}$ versus a similarly scaled 
pseudoscalar meson mass (calculated, of course, with valence 
quarks that are degenerate with those in the sea, i.e.
$\kappa_{\rm valence} = \kappa_{\rm sea}$).  For each point we also
plot the corresponding value of the quenched topological
susceptibility, calculated at the same lattice spacing
using the interpolation formula
Eqn.~\ref{eqn_chiqu_int}. It is useful to note at this point the small
variation in the quenched values over the range of data points
simulated. This is a consequence of the UKQCD strategy of matching
$\hat{r}_0$. Without this, if the calculations were performed
at fixed $\beta$, the lattice spacing would become
increasingly coarse with increasing $m_q$, and the reduction in 
the quenched susceptibility would be much more pronounced over this 
range of $\kappa$. We shall return to this important point 
when we discuss other work in a later section.

Comparing the dynamical and quenched values, the
effects of the sea quarks are clear. Whilst the $(5.29,0.1340)$ point
is consistent with the quenched value, moving to smaller $m_q$
($\propto m_\pi^2$) the topological susceptibility is increasingly
suppressed.  

\begin{figure}[tb]
\begin{center}
\leavevmode
\epsfysize=510pt
\epsfbox[20 30 620 730]{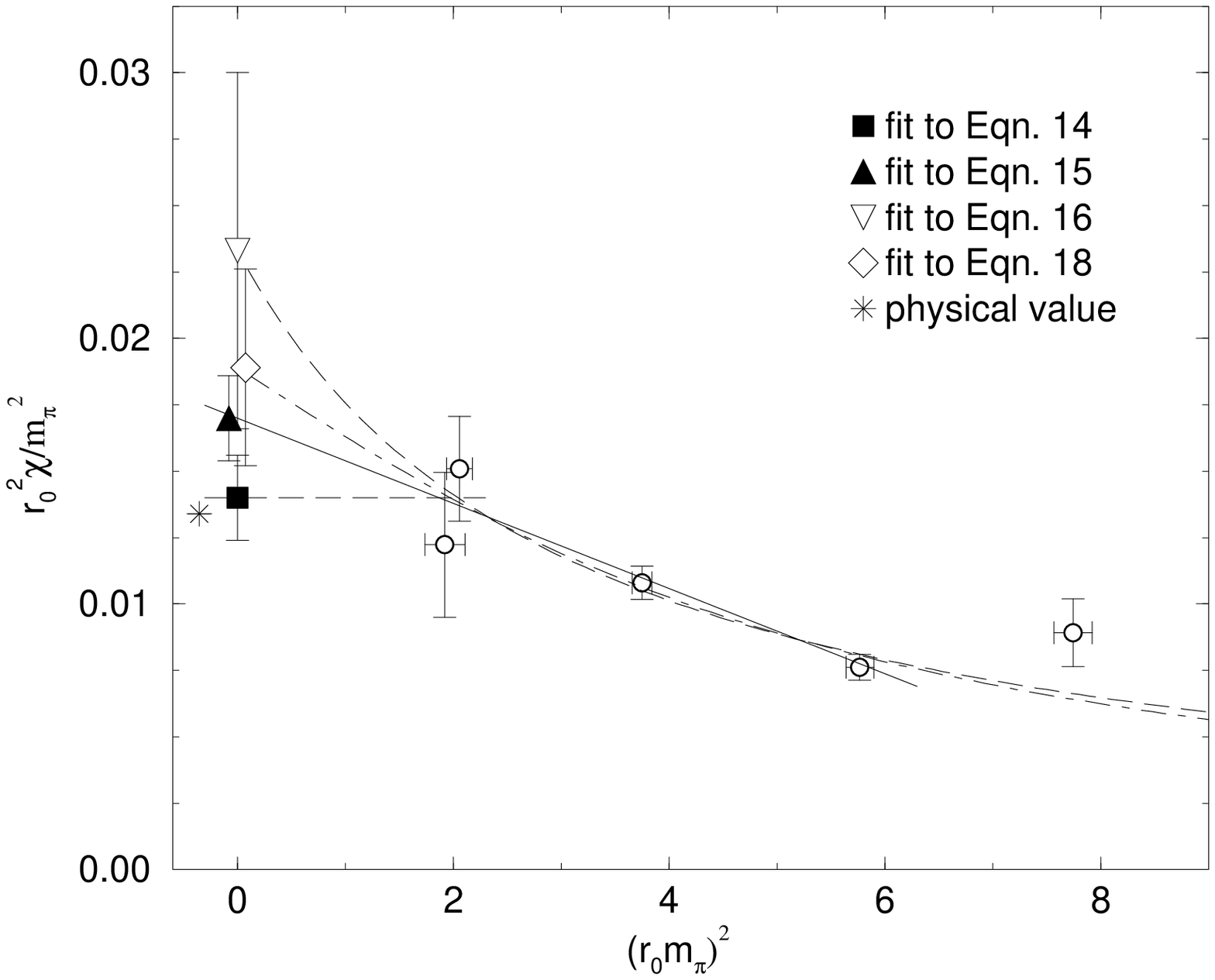}
\end{center}
\vspace{-1.0cm}
\caption[]{\label{fig_r02}
  {\it Fits to the measured topological susceptibility, independent of
    the quenched points.}}
\end{figure}

We can make this observation more quantitative by attempting to fit
our values of $\hat{r}_0^4\hat{\chi}$ with the expected functional 
form in Eqn.~\ref{eqn_chi_pi2}. 
But we must first be clear whether this fit is justified, and what
exactly we are extrapolating in.  Eqn.~\ref{eqn_chi_pi2} is strictly a
chiral expansion that describes the behaviour for small sea quark
masses in the continuum and relates it to $f_\pi$. We expect that 
it is also applicable for a set of lattice data points evaluated at
the same lattice spacing, except that now the decay constant,
$\hat{f}_\pi$, will be the one appropriate to that lattice spacing. 
Whilst most of our data points are evaluated on a trajectory 
of constant lattice spacing in the parameter space, 
the two most chiral values are not. The lattice spacings there are 
slightly finer. If $\hat{r}_0 \hat{f}_\pi$ varied significantly with
$a$ over this range of $a$, it would not be clear how to perform
a consistent chiral extrapolation through the data points.
The non--perturbative improvement
of the action, however, removes the leading order lattice spacing
dependence and the residual corrections in this range of lattice
spacings appear to be small, at least in measurements of hadron
spectroscopy
\cite{wittig98,edwards98}.
Assuming the variation in the decay constants is similarly small
over what, it should be recalled, is not a large variation in $a$, we
may proceed to attempt a common chiral extrapolation to the data. For
this purpose it is useful to redisplay the data in
Fig.~\ref{fig_r02}, where the leading order chiral behaviour would
then be a horizontal line, 
\be
\frac{\hat{r}_0^2\hat{\chi}}{\hat{m}_\pi^2} =
c_0 
\label{eqn_fit_fl} \\
\ee
and including the leading correction gives a general straight line
\be
\frac{\hat{r}_0^2\hat{\chi}}{\hat{m}_\pi^2}
 = 
c_0 + c_1 (\hat{r}_0 \hat{m}_\pi)^2.
\label{eqn_fit_fl_lin} 
\ee
In each case the intercept is related to the decay constant by $c_0 =
(\hat{r}_0 \hat{f}_\pi)^2/4$. We now follow a standard fitting 
procedure, first using the most chiral points, then systematically
adding the  less chiral points until the fit becomes unacceptably 
bad. The larger the number of points one can add in this way, the 
more evidence one has for the fitted form and the more
confident one is that the systematic errors, associated with
the neglected higher order corrections, are small.
The results of performing such fits are shown in
Table~\ref{tab_fit_res} and those using the two and four most 
chiral points respectively are plotted on Fig.~\ref{fig_r02}.
We see from the Table that the fits using Eqn.~\ref{eqn_fit_fl_lin}
show much greater stability and these are the ones that will
provide our eventual best estimate for $f_\pi$.

We should comment briefly on the determination of the fitting
parameter errors.  In performing all but the flat fit we must contend
with the data having (small) errors on the abscissa in addition to the
ordinate. In order to estimate their affect on the fitting parameters,
we first perform fits to the data assuming that the abscissa data take
their central values. Identical fits are then made using the central
values plus one, and then minus one standard deviation. The spread of
the fit parameters obtained provides what is probably a crude
over--estimate of this error (given there is some correlation between
the ordinal and abscissal uncertainties) but is sufficient to show
that it is minor. We show this spread as a second error, and
for estimates of the decay constant we add it in quadrature to the
other fit parameter error.

It is remarkable that we can obtain stable fits to most of our data 
using just the first correction term in Eqn.~\ref{eqn_fit_fl_lin}.
Nonetheless, as we can see in Fig.~\ref{fig_r04}, our values
of the susceptibility are not very much smaller than the 
$m_\pi = \infty$ quenched value and we need to have some
estimate of the possible systematic errors that may
arise from neglecting
the higher order corrections that will eventually check
the rise in $\hat{\chi}$. As discussed earlier we shall do so
by exploring two possibilities. One is that the reason 
why  $\hat{\chi}$ is close to $\chiquhat$ is not that
$m_q$ is `large' but rather that $N_c = 3$ is large.
Then the values of $\hat{\chi}$ should follow the form in
Eqn.~\ref{eqn_nlge_form}. A second possibility is simply
that our values of $m_q$ are indeed large. In that case
we have argued that the functional form Eqn.~\ref{eqn_mlge_form}
should be a reasonable representation of the true mass
dependence. We now perform  both types of fit in turn. 

We begin with the first possibility, and therefore fit the data
with the following ansatz
\be
\hat{r_0}^4 \hat{\chi} =  
\frac{c_0 c_3 (\hat{r}_0 \hat{m}_\pi)^2}
{c_3 + c_0 (\hat{r}_0 \hat{m}_\pi)^2},
\label{eqn_fit_nlge}
\ee
where we expect $c_3$ = $\hat{r_0}^4 \chiquhat$ up to ${\cal
  O}(1/N_c^2)$ corrections.  To test this we fit up to six data
points.  The first five are measured in the dynamical simulations. The
final quantity is the quenched susceptibility at a lattice spacing
equivalent to those of the three least chiral, ``matched'' ensembles.
The last is assigned an arbitary large value $(\hat{r}_0
\hat{m}_\pi)^2 = 1000$ whose exact value has no effect on the fits.

We also expect, from the Maclaurin chiral expansion of
Eqn.~\ref{eqn_fit_nlge}, \be \hat{r_0}^4 \hat{\chi} = c_0 \cdot
(\hat{r}_0 \hat{m}_\pi)^2 - c_0^2/c_3 \cdot (\hat{r}_0 \hat{m}_\pi)^4
+ c_0^3/c_3^2 \cdot (\hat{r}_0 \hat{m}_\pi)^6 + {\cal O}((\hat{r}_0
\hat{m}_\pi)^8)
\label{eqn_exp_nlge}
\ee
that $c_0$ is related to the decay constant as before, $c_0 =
(\hat{r}_0 \hat{f}_\pi)^2/4$. We present the results of the
fits in Table~\ref{tab_fit_res2}.

We turn now to fits based on the functional form in
Eqn.~\ref{eqn_mlge_form}. We therefore use the ansatz 
\be
\hat{r_0}^4 \hat{\chi}
= 
\frac{2c_0}{\pi} (\hat{r}_0 \hat{m}_\pi)^{2} 
\tan^{-1} \left(
  \frac{c_3}{\frac{2c_0}{\pi}(\hat{r}_0 \hat{m}_\pi)^{2}} \right)
\label{eqn_fit_atan} 
\ee
where once again we expect $c_3$ = $\hat{r_0}^4 \chiquhat$ and
from the expansion
\be
\hat{r_0}^4 \hat{\chi}
 = 
c_0 (\hat{r}_0 \hat{m}_\pi)^2
- \left( \frac{2c_0}{\pi} \right) ^2 \cdot 
\left( \frac{1}{c_3} \right) \cdot (\hat{r}_0 \hat{m}_\pi)^4 
+ {\cal O}((\hat{r}_0 \hat{m}_\pi)^8)
\label{eqn_exp_atan} 
\ee
we expect  $c_0 = (\hat{r}_0 \hat{f}_\pi)^2/4$. Note that in
contrast to Eqn.~\ref{eqn_exp_nlge}, Eqn.~\ref{eqn_exp_atan}
has no term that is cubic in $m_q$ and the rise will remain 
approximately quadratic for a greater range in
$(\hat{r}_0 \hat{m}_\pi)^2$. That this need be no bad thing
is suggested by the relatively large range over which we could fit
Eqn.~\ref{eqn_fit_fl_lin}. And indeed we see from the fits listed
in Table~\ref{tab_fit_res2} that this form fits our data quite 
well.

Typical examples of the fits from Eqn.~\ref{eqn_fit_nlge}
and Eqn.~\ref{eqn_fit_atan} 
are shown in Figs.~\ref{fig_r04} and~\ref{fig_r02}. The similarity of
the two functions is apparent. In Table~\ref{tab_fit_chir} we use the
fit parameters to construct the first three expansion coefficients in
the Maclaurin series for the various fit functions, describing the
chiral behaviour of $\chi$. The fits are consistent with one another.

The fitted asymptote of the susceptibility at large $m_q$ is given by
$c_3$. We see from Table~\ref{tab_fit_res2} that these are broadly
consistent with the quenched value, and our large statistical errors
do not currently allow us to resolve any ${\cal O}(1/N_c^2)$
deviation from this.

As an aside we ask what happens if we cast aside some of our 
theoretical expectations and ask how strong is the evidence 
from our data that (a) the dependence is on $m_\pi^2$ rather than 
on some other power, and (b) the susceptibility really does go to
zero as $m_\pi \to 0$? To answer the first question
we perform fits of the kind in   Eqn.~\ref{eqn_fit_atan}
but replacing $(\hat{r}_0 \hat{m}_\pi)^{2}$ by
$(\hat{r}_0 \hat{m}_\pi)^{c}$. We find, using all six values of
$\hat{\chi}$, that $c = 1.64 (22) (8)$; a value consistent
with $c = 2$. The $\chi^2$/d.o.f. is poorer, however, than
for the fit with a power fixed to 2 (possible as 
there is one fewer d.o.f.)
suggesting that the data does not warrant the use of such an 
extra parameter. As for the second question, we add a constant
$\hat{c}$ to  Eqn.~\ref{eqn_fit_atan} and find 
$\hat{c} = 0.0096 (78) (20)$. Again this is consistent with our
theoretical expectation; and again the $\chi^2$/d.o.f. is worse.
(See Table~\ref{tab_fit_res3} for details of the above two fits.)

Given the consistency of our description of the small $m_\pi$ regime
from our measurements it is reasonable to use the values of $c_0$ to
estimate the pion decay constant, $f_\pi$. This is done in units of
$r_0$ in Tables~\ref{tab_fit_res} and~\ref{tab_fit_res2}. 
We use the chiral fit of Eqn.~\ref{eqn_fit_fl_lin} over the 
largest acceptable range to provide us with our best estimate
and its statistical error. We then use the fits with other functional
forms to provide us with the systematic error. This produces
an estimate
\be
\hat{r}_0 \hat{f}_\pi  =  0.261 \ \pm 0.013 \ ^{+0.045}_{-0.025}
\label{eqn_r0fpi} 
\ee
where the first error is statistical and the second is systematic.
This is of course no more than our best estimate of the value of
$f_\pi$ corresponding to our lattice spacing of $a \simeq 0.1 \fm$.
This value will contain corresponding lattice spacing corrections and
these must be estimated before making a serious comparison with the
experimental value.  Our use of the non-perturbative $c_{\rm sw}$
should have eliminated the leading ${\cal O}(a)$ errors, however, and
$(a/r_0)^2$ is sufficiently small that it is plausible to expect these
lattice corrections to be smaller than our other errors.  If we use
the phenomenological value $r_0 = 0.49 \ \fm$ then we obtain from
Eqn.~\ref{eqn_r0fpi} the value
\be
 f_\pi  =  105 \ \pm 5 \ ^{+18}_{-10} \ \MeV
\label{eqn_mevfpi} 
\ee
which is reasonably close to the experimental value
$\simeq 93 \ \MeV$. In 
\cite{ukqcd_prog} 
a comparison with other physical quantities is expected to
provide us with a quantitative estimate of the magnitude
of the ${\cal O}(a^2)$ errors.

\section{Comparison with other studies}
\label{sec_compare}

During the course of this work, there have appeared some other 
studies of the topological susceptibility in lattice QCD; in 
particular one by the CP-PACS collaboration
\cite{cppacs99},
and one by the Pisa group
\cite{alles99}.
Both calculations find no significant decrease of the susceptibility 
with decreasing quark (or pion) mass when everything is expressed 
in physical, rather than lattice, units. This appears to contradict 
our findings and is clearly something we need to address. 

Our suggestion why these other studies have seen no decrease in 
$\chi$ is as follows. They differ from our study in having been
performed at fixed $\beta$. That implies that the lattice
spacing $a$ decreases as $m_q$ is decreased. In typical current
calculations this variation in $a$ is substantial. (See for
example Fig.~4 of 
\cite{ukqcd98}.)
At the smallest values of $m_q$ the lattice spacing cannot be allowed
to be too fine, because the total spatial volume must remain
adequately large. This implies that at the larger values of $m_q$ the
lattice spacing is quite coarse.  Over such a range of lattice
spacings the topological susceptibility in the pure gauge theory
typically shows a large variation. Since for coarser $a$ more
instantons (those with $\rho \leq {\cal O}(a)$) are excluded, and more
of those remaining are narrow in lattice units (with a correspondingly
suppressed lattice topological charge) we expect that this variation
is quite general, and not a special feature of the pure gauge theory
with a Wilson action. In lattice QCD we therefore expect two
simultaneous effects in $\hat{\chi}$ as we decrease $m_q$ at fixed
$\beta$.  First, because of the ${\cal O}(a^2)$ lattice corrections
just discussed, $\hat{\chi}$ will, like $\chiquhat$, tend to increase.
Second, it will tend to decrease because of the physical quark mass
dependence.  In the range of quark masses covered in current
calculations this latter decrease is not very large (as we have seen
in our work) and we suggest that the two effects may largely
compensate each other so as to produce a susceptibility that shows
very little variation with $m_q$; in contrast to the ratio
$\hat{\chi}/\chiquhat$ which does.
 
To illustrate this consider the fixed-$\beta$ calculation in
\cite{ukqcd98}.
The range of quark masses covered in that work corresponds to 
$(\hat{r}_0 \hat{m}_\pi)^2$  decreasing from about 6.5 to about 3.0.
Simultaneously $a/r_0$ decreases from about 0.437 to about 0.274.
Over this range of $1/\hat{r}_0 \equiv a/r_0$ the 
pure gauge susceptibility increases by almost a factor of two,
as we see using Eqn.~\ref{eqn_chiqu_int}. Clearly this
is large enough to compensate for the hoped--for near--chiral
variation of the susceptibility.

To make this a little more quantitative we consider the different 
calculations in turn, beginning with that in
\cite{alles99}.
The Pisa group uses the Wilson gauge action, and four flavours of
staggered fermions. Simulations are carried out at a fixed value of
the gauge coupling for a range of bare quark masses between $\hat{m}_q
= 0.01$ and 0.05.  Expressed in units of the string
tension, $\sigma$, this corresponds to a change in the quark mass
by a factor of approximately three. Over the corresponding
range of lattice spacings the susceptibility in the pure 
gauge theory with Wilson action varies by a factor of $1.44 (19)$, 
when determined using the same ``field theoretic + smearing'' 
method (for a review, see
\cite{teper98b}).
Such an increase would be large enough to largely compensate for the
hoped--for near chiral decrease of the susceptibility (see for example
the variation of $\hat{r}^4_0 \hat{\chi}$ between $(\hat{r}_0
\hat{m}_\pi)^2 \simeq 6.0$ and $\simeq 2.0$ in Fig.~\ref{fig_r04}).

The CP-PACS study is carried out using an improved (Iwasaki) gauge
action with two flavours of tadpole--improved Sheikoslami--Wohlert
clover fermions. Combining the data from
\cite{cppacs99,cppacs},
one finds over the range of quark masses considered a variation in the
square of the pion mass (in units of the string tension) of
approximately 3, similar to that of the Pisa study. Over this range
there is also little variation in the topological susceptibility when
expressed in units of the string tension. The string tensions
considered would correspond, in a pure gauge Wilson action study, to
a range in $\beta$ of $5.64-5.71$, and in that range the topological
susceptibility, $\chiquhat$, varies by a factor of around 1.5.
If we assume that there is a similar effect with the Iwasaki action,
then over this range the  dynamical $\hat{\chi}$ appears to suffer
a suppression of about a factor of 1.5 relative to the
pure gauge theory. As before, not all the
data is near the chiral limit, and over a corresponding range in the
UKQCD data we see a suppression of around a factor of 2. CP-PACS
in fact possess datasets at four values of $\beta$ and they
have calculated the topological susceptibility on the dataset
corresponding to the second smallest value. At their higher values
of $\beta$ the effect we have been describing here will be
much weaker and we believe that calculations performed on those
two datasets will reveal a quark mass dependence comparable to
the one we find.

The arguments presented in this section are by no means definitive. We
do not, for instance, know the discretisation effects on the
topological susceptibility in lattice gluodynamics with the Iwasaki
action. Neither do we know know how the (different) fermionic
actions alter these discretisation effects. What
the heuristic analysis given here aims to show is that these
results are not necessarily in contradiction to the theoretical
expectations and to the results presented in this talk.

\section{Conclusions}
\label{sec_conc}

We have calculated the topological susceptibility in lattice QCD
with two light quark flavours, using lattice field configurations,
recently generated by UKQCD, in which the lattice spacing
is approximately constant as the quark mass is varied.
We find that there is clear evidence of the effects of sea quarks 
in suppressing $\chi$.

We discuss this behaviour in the context of chiral and large $N_c$
expansions, and find good agreement with the functional forms
expected there. We are not able to make a stronger statement
about how close QCD is to its large $N_c$ limit, owing to the 
relatively large statistical errors on our calculated values,
particularly at larger quark masses. 

The consistent leading order chiral behaviour from our various fitting
ans\"{a}tze allows us to make an estimate for the pion decay constant,
$f_\pi = 105 \ \pm 5 \ ^{+18}_{-10} \ \MeV$, for the lattice spacing of
$a \simeq 0.1 \fm$. (Here the first error is statistical and the
second has to do with the chiral extrapolation.)  Since we use a
lattice fermion action in which the leading ${\cal O}(a)$
discretisation errors have been removed, and (quenched) hadron masses
show little residual lattice spacing dependence, we might expect that
this value is close to its continuum limit. In any case we note that
it is in agreement with the experimental value, $\simeq 93 \ \MeV$.

We have commented on some recent studies of the topological 
susceptibility in lattice QCD, which do not see the kind of mass 
dependence that we claim to have observed. We have suggested 
that this might arise from the fact that in these fixed-$\beta$ 
calculations the value of $a$ decreases with decreasing $m_\pi^2$
and the consequent variation in those ${\cal O}(a^2)$ lattice corrections
that the susceptibility (plausibly) shares with the pure gauge theory
might be large enough to compensate for the expected near--chiral
variation. Further analysis of this and the
corresponding quenched theories in this light would be welcome.

\section*{Acknowledgments}

The work of A.H. was supported in part by UK PPARC grant
PPA/G/0/1998/00621. A.H. wishes to thank the Aspen Center for Physics
for its hospitality during part of this work.

\newpage

\begin{table}[tb]
\begin{center}
\begin{tabular}{|c|r@{.}l|}
\hline
\hline
$(\beta,\kappa)$ & \multicolumn{2}{c|}{$\hat{\chi} \times 10^5$} \\
\hline
(5.20, 0.13565) & 3&18 (64) \\
(5.20, 0.13550) & 4&84 (57) \\
(5.20, 0.13500) & 7&90 (42) \\
(5.26, 0.13450) & 8&86 (54) \\
(5.29, 0.13400) & 12&9 (1.9) \\
\hline
\hline
\end{tabular}
\caption{ \label{tab_chi}
  {\em The topological susceptibility in lattice units.}}
\end{center}
\end{table}
\begin{table}[tb]
\begin{center}
\begin{tabular}{|c|c|r@{.}l|r@{.}l|r@{.}l|r@{.}l|}
\hline
\hline
Fit & 
$N_{\rm fit}$ &
\multicolumn{2}{c|}{$c_0$} &
\multicolumn{2}{c|}{$c_1$} &
\multicolumn{2}{c|}{$\chi^2/{\rm d.o.f.}$} &
\multicolumn{2}{c|}{$\hat{r}_0 \hat{f}_\pi$} \\
\hline
Eqn.~\ref{eqn_fit_fl} & 2 &  
0&0140 (16)  & \none & 0&805 &  0&237 (14) \\
Eqn.~\ref{eqn_fit_fl} & 3 &  
0&0112 (6) & \none & 2&202 &  0&212 (6)  \\
Eqn.~\ref{eqn_fit_fl} & 4 &  
0&0091 (4) & \none & 9&008 & \none \\
\hline
Eqn.~\ref{eqn_fit_fl_lin} & 3  & 
0&0176 (35) (4)  & $-0$&0018 (10) (1) & 0&964 & 0&265 (27) \\
Eqn.~\ref{eqn_fit_fl_lin} & 4  & 
0&0170 (16) (1)  & $-0$&0016 (4) (0) & 0&502  & 0&261 (13) \\
Eqn.~\ref{eqn_fit_fl_lin} & 5  & 
0&0147 (14) (1)  & $-0$&0011 (3) (0) & 2&965 & 0&242 (12) \\
\hline
\hline
\end{tabular}
\caption{ \label{tab_fit_res}
  {\em Fits to the $N_{\rm fit}$ most chiral points of
    $(\hat{r}_0^2\hat{\chi})/\hat{m}_\pi^2$.}}
\end{center}
\end{table}
\begin{table}[tb]
\begin{center}
\begin{tabular}{|c|c|r@{.}l|r@{.}l|r@{.}l|r@{.}l|}
\hline
\hline
Fit & 
$N_{\rm fit}$ &
\multicolumn{2}{c|}{$c_0$} &
\multicolumn{2}{c|}{$c_3$} &
\multicolumn{2}{c|}{$\chi^2/{\rm d.o.f.}$} &
\multicolumn{2}{c|}{$\hat{r}_0 \hat{f}_\pi$} \\
\hline
Eqn.~\ref{eqn_fit_nlge} & 3 & 
0&0208 (87) (12) & 0&0844 (427) (35) &  
1&013 & 0&288 (61) \\
Eqn.~\ref{eqn_fit_nlge} & 4 & 
0&0272 (85) (18) & 0&0632 (114) (6) &  
0&895 & 0&329 (53) \\
Eqn.~\ref{eqn_fit_nlge} & 5 & 
0&0233 (66) (10) & 0&0717 (147) (3) &  
1&847 &  0&305 (44) \\
Eqn.~\ref{eqn_fit_nlge} & 6 & 
0&0279 (32) (11) & 0&0631 (5) (0) &  
1&489 &  0&334 (21) \\
\hline
Eqn.~\ref{eqn_fit_atan} & 3 & 
0&0186 (53) (7) & 0&0576 (175) (6) &  
0&990 & 0&273 (40) \\
Eqn.~\ref{eqn_fit_atan} & 4 & 
0&0209 (42) (7) & 0&0506 (55) (5) &  
0&682 & 0&289 (30) \\
Eqn.~\ref{eqn_fit_atan} & 5 & 
0&0189 (36) (5) & 0&0550 (69) (6) &  
1&929 &  0&275 (27) \\
Eqn.~\ref{eqn_fit_atan} & 6 & 
0&0164 (14) (7) & 0&0629 (5) (0) & 
1&567 &  0&256 (12) \\
\hline
\hline
\end{tabular}
\caption{ \label{tab_fit_res2}
  {\em Fits to the $N_{\rm fit}$ most chiral points of
    $(\hat{r}_0^4\hat{\chi})$.}}
\end{center}
\end{table}
\begin{table}[tb]
\begin{center}
\begin{tabular}{|c|c|r@{.}l|r@{.}l|r@{.}l|}
\hline
\hline
Fit & 
$N_{\rm fit}$ &
\multicolumn{2}{c|}{const.} &
\multicolumn{2}{c|}{${\cal O}((\hat{r}_0\hat{m}_\pi)^2)$} &
\multicolumn{2}{c|}{${\cal O}((\hat{r}_0\hat{m}_\pi)^4)$} \\
\hline
Eqn.~\ref{eqn_fit_fl_lin} & 4 &
0&0170 (17) & $-0$&0016 (4) & \multicolumn{2}{c|}{---}  \\
\hline
Eqn.~\ref{eqn_fit_nlge} & 5 &
0&0233 (67) & $-0$&0031 (19) & 0&0025 (24) \\
Eqn.~\ref{eqn_fit_nlge} & 6 &
0&0279 (34) & $-0$&0123 (30) & 0&0055 (19) \\
\hline
Eqn.~\ref{eqn_fit_atan} & 5 &
0&0189 (37) & $-0$&0026 (11) & \multicolumn{2}{l|}{0}  \\
Eqn.~\ref{eqn_fit_atan} & 6 &
0&0164 (16) & $-0$&0017 (4) & \multicolumn{2}{l|}{0}  \\
\hline
\hline
\end{tabular}
\caption{ \label{tab_fit_chir}
{\em Chiral expansion terms of fitted functions.}}
\end{center}
\end{table}
\begin{table}[tb]
\begin{center}
\begin{tabular}{|c|r@{.}l|r@{.}l|r@{.}l|r@{.}l|r@{.}l|}
\hline
\hline
$N_{\rm fit}$ &
\multicolumn{2}{c|}{$c_0$} &
\multicolumn{2}{c|}{$c_3$} &
\multicolumn{2}{c|}{$c$} &
\multicolumn{2}{c|}{$\hat{c}$} &
\multicolumn{2}{c|}{$\chi^2/{\rm d.o.f.}$} \\
\hline
6 & 
0&0112 (40) (5) & 0&0533 (78) (17) &  \multicolumn{2}{c|}{\rm fixed to 2}  & 
0&0096 (78) (20) &
1&825 \\
6 & 
0&0206 (26) (16) & 0&0631 (80) (49) & 1&64 (22) (8)  &  
\multicolumn{2}{c|}{\rm fixed to 0}  &
1&798 \\
\hline
\hline
\end{tabular}
\caption{ \label{tab_fit_res3}
{\em Fits to the $N_{\rm fit}$ most chiral values  of
$(\hat{r}_0^4\hat{\chi})$;
for the power $c$ of $m^2_\pi$, and for the chiral intercept,
$\hat{c}$. These are as described in the text.}}
\end{center}
\end{table}


\begin{thebibliography}{99}

\bibitem{teper99}
M. Teper,
hep-lat/9909124.

\bibitem{hart99}
A. Hart, M. Teper, 
hep-lat/9909072.

\bibitem{ukqcd_prog}
UKQCD collaboration, 
in preparation.

\bibitem{vecchia80}
P. Di Vecchia, G. Veneziano, 
Nucl. Phys. B 171 (1980) 253.

\bibitem{leutwyler92}
H. Leutwyler, A. Smilga,
Phys. Rev. D 46 (1992) 5607.

\bibitem{thooft74}
G. `t Hooft,
Nucl. Phys. B72 (1974) 461.

\bibitem{witten79}
E. Witten,
Nucl. Phys. B160 (1979) 57.

\bibitem{teper98a}
M. Teper,
Phys. Rev. D 59 (1999) 014512
[hep-lat/9804008].

\bibitem{teper98b}
M. Teper,
hep-th/9812187.

\bibitem{ukqcd99}
J. Garden (UKQCD),
hep-lat/9909066.

\bibitem{irving98}
A. Irving et al.,
Phys. Rev. D 58 (1998) 114504
[hep-lat/9807015].

\bibitem{sommer94}
R. Sommer,
Nucl. Phys. B411 (1994) 839
[hep-lat/9310022].

\bibitem{smith98}
D. Smith, M. Teper,
Phys. Rev. D 58 (1998) 014505
[hep-lat/9801008].

\bibitem{divecchia81}
P. di Vecchia et al.,
Nucl. Phys. B192 (1981) 392.

\bibitem{pisa88}
M. Campostrini, A. Di Giacomo, H. Panagopoulos,
Phys. Lett. B212 (1988) 206.

\bibitem{wittig98}
H. Wittig,
Nucl. Phys. (Proc.Suppl.) 63 (1998) 47
[hep-lat/9710013].

\bibitem{edwards98}
R. Edwards, U. Heller, T. Klassen,
Phys. Rev. Lett. 80 (1998) 3448
[hep-lat/9711052].

\bibitem{cppacs99}
A. Ali Khan et al. (CP-PACS),
hep-lat/9909045.

\bibitem{alles99}
B. All\'es, M. D'Elia, A. Di Giacomo,
hep-lat/9912012.

\bibitem{ukqcd98}
C. Allton et al. (UKQCD),
Phys.Rev. D60 (1999) 034507,
[hep-lat/9808016].

\bibitem{cppacs}
A. Ali Khan et al. (CP-PACS),
hep-lat/9909050; 
S. Aoki et al. (CP-PACS), 
hep-lat/9809120;
R. Burkhalter,
private communication.

\end{thebibliography}
\end{document}